\documentclass[%
 reprint,
 superscriptaddress,
 amsmath,amssymb,
 aps,
]{revtex4-2}

\usepackage{graphicx}% Include figure files
\usepackage{dcolumn}% Align table columns on decimal point
\usepackage{bm}% bold math
\begin{document}

%\preprint{APS/123-QED}

%%%%%
\title{Finite element method modeling of expansion of irradiated rocks: focusing on the minerals
%Prediction of expansion of irradiated aggregate by using finite element method
}% Force line breaks with \\

%%%%%

\author{Y. Oida}
\affiliation{MRI Research Associates, Inc., Chiyoda, Tokyo 100-0014 Japan.}
\author{S. Sakuragi}
\email{shunsuke\_sakuragi@mri-ra.co.jp}
\affiliation{MRI Research Associates, Inc., Chiyoda, Tokyo 100-0014 Japan.}
\author{T. Igari}
\affiliation{MRI Research Associates, Inc., Chiyoda, Tokyo 100-0014 Japan.}
\author{Y. Nakajima}
\affiliation{MRI Research Associates, Inc., Chiyoda, Tokyo 100-0014 Japan.}
\author{E. Ro}
\affiliation{MRI Research Associates, Inc., Chiyoda, Tokyo 100-0014 Japan.}
\author{F. Shimizu}
\affiliation{MRI Research Associates, Inc., Chiyoda, Tokyo 100-0014 Japan.}
\author{M. Futami}
\affiliation{MRI Research Associates, Inc., Chiyoda, Tokyo 100-0014 Japan.}
\author{Y. Hakozaki}
\affiliation{Graduate School of Engineering, Chiba University, Inage, Chiba 263-8522, Japan.}
\author{T. Ohkubo}
\affiliation{Graduate School of Engineering, Chiba University, Inage, Chiba 263-8522, Japan.}
\author{H. Ishikawa}
\affiliation{Department of Mechanical Systems Engineering, Tokyo University of Agriculture and Technology, Koganei, Tokyo 184–8588, Japan.}
\author{S. Takada}
\affiliation{Department of Mechanical Systems Engineering, Tokyo University of Agriculture and Technology, Koganei, Tokyo 184–8588, Japan.}
\affiliation{Institute of Engineering, Tokyo University of Agriculture and Technology, Koganei, Tokyo 184–8588, Japan.}
\author{S. Sawada}
\affiliation{Kajima Corporation, Minato, Tokyo 107-8388 Japan.}
%\author{O. Kontani}
%\affiliation{Kajima Corporation, Minato, Tokyo 107-8388 Japan.}
\author{K. Murakami}
\affiliation{Graduate School of Engineering, The University of Tokyo, Bunkyo, Tokyo 113-8656, Japan.}
\author{K. Suzuki}
\affiliation{Mitsubishi Research Institute, Inc., Chiyoda, Tokyo 100-0014 Japan.}
\author{I. Maruyama}
\email{i.maruyama@bme.arch.t.u-tokyo.ac.jp}
\affiliation{Graduate School of Engineering, The University of Tokyo, Bunkyo, Tokyo 113-8656, Japan.}
\affiliation{Graduate School of Environmental Studies, Nagoya University, Chikusa, Nagoya 464-8603, Japan.}

\date{\today}

\begin{abstract}
Finite element method (FEM) modeling of the volumetric expansion phenomenon associated with the accumulation of irradiation was performed on rocks in a concrete for nuclear power plant. 
The FEM mesh of sandstone, tuff, and granite was created based on microscopic images, and the volumetric expansion due to irradiation was calculated. The saturated value of the volume expansion due to irradiation accumulation was calculated, and it was shown that the saturated value of the volume expansion was well explained by the experimental value.
In addition, the FEM analysis indicates that irradiation-induced degradation of rock propagates through a localized concentration of stress followed by fracture with cracks spreading throughout the rock. 

\end{abstract}

\maketitle

%\tableofcontents
\section{Introduction}
In Japan, nuclear power plants have been in continuous operation for 30 years, and there is a need for a method to evaluate the deterioration of power plant buildings from the viewpoint of safety. 
Since nuclear power plant buildings are concrete structures, the safety of the buildings will be determined by evaluating the deterioration of the concrete, as is the case with ordinary buildings. 
On the other hand, concrete used in nuclear power plants is under a special environment where it is constantly exposed to neutrons for several decades, and thus, the degradation phenomena of general structures cannot be appropriated and discussed in the same way. Therefore, there is a need to clarify the mechanism of degradation phenomena of concrete caused by neutron irradiation in order to obtain general knowledge on the safety of nuclear power plants  \cite{braverman2004degradation, le2020irradiation, maruyama2021post}.

In previous reports, irradiation degradation of concrete has been discussed from an experimental point of view: expansion and degradation due to irradiation accumulation in the rocks configuring the concrete \cite{hilsdorf1978effects, maruyama2018impact, YannLePape2018}. 
Dimensional measurements have revealed the existence of irradiation-induced expansion from a macroscopic viewpoint, and it has been reported that volume expansion of up to 20\% or more occur in sandstone, tuff, and granite \cite{Denisov}.
In addition, x-ray diffraction (XRD) measurement on rocks have also discussed degradation phenomena from a microscopic viewpoint, such as changes in the lattice constants of mineral crystals constituting the rocks, and they reveals that these irradiation expansions are caused by the amorphization of mineral crystals associated with neutron irradiation \cite{Denisov, devine1994macroscopic, devine1990physical, hobbs1995role}. 
On the other hand, even for quartz with high radiosensitivity, the volume expansion obtained from XRD is about $\sim$18\%, which cannot explain the macroscopic results based on dimensional measurements. 
This phenomenon is qualitatively understood to be the expansion of the mineral crystals constituting the rocks, which causes cracks throughout the rocks and thus expands the rocks beyond the effect of amorphization. 
However verifying this phenomenon experimentally, experiments using neutron reactors have not provided systematic experimental data due to the difficulty of conducting systematic experiments on a large number of rock samples, and thereby a universal mechanism explaining the irradiation swelling of rock from the mineral perspective has not been clarified.

In contrast to clarifying the universal curve of irradiation degradation in rocks depending on the constituent mineral species, an approach utilizing computer science is useful for systematically discussing the relationship between the expansion of various minerals and rock degradation. 
Therefore, as a preparatory step, this study examined a method to create an FEM model focusing on the minerals constituting the rock from photographs of the actual rock. 
The FEM model considering the fracture of grain boundaries for sandstone, tuff, and granite reproduced well the maximum value of irradiation expansion obtained from the experiment. We propose that the FEM model can be used to understand the irradiation degradation and expansion of rockss using computational methods.

\section{modelling and method}
Ansys LS-DYNA, a nonlinear structural analysis software developed by Livermore Software Technology Corporation, was used for the FEM analysis \cite{Dyna}.
Since rock fracture is well explained by Griffith theory \cite{griffith1921vi, liebowitz1968mathematical}, the FEM constitutive equation (spring-damper model: MAT189 in LS-DYNA) was used, and modeling was performed considering fracture at the grain boundary. The expansion process associated with the accumulation of irradiation was treated as thermal expansion reaching thermal equilibrium and calculated by FEM.
FEM mesh elements that were determined to be destructive were deleted to express the destruction.

Polarized light microscope images were taken of sandstone, tuff, and granite surfaces to create the FEM mesh. The mineral species were identified by image analysis using color resolved \cite{chen2004digital} watershed algorithm, and voxel meshes were generated \{Fig. 1 (a) and (b)\}. 
The percentages of each mineral in the rock obtained from image analysis are shown in Table 1. 
Note that this value agrees well with the value obtained from the XRD measurement within an error of $\sim$1\%. 
For the irradiation dependence of coefficient of linear expansion, values reported in Denisov's report were used \cite{Denisov}. The physical properties of the minerals were obtained from molecular dynamics (MD) calculations \cite{okada2020characterization, Ohkubo}. Since the crystallographic orientation of the minerals cannot be determined from polarized light microscope images, the coefficients of linear expansion were set by randomly assigning crystallographic orientations to the minerals in this model. 

And then, fracture stresses were determined so that the FEM Brazillian test reproduced the tensile strength values obtained from reports of previous experiments. 
For FEM analysis of the Brazilian test, a circularly cut mesh of polarized light microscope images \cite{sawada} was used. The lower end of the circle was constrained and forced displacement was applied from the upper end to simulate the Brazilian test. For simplicity, the vertical and shear failure stresses were assumed to be the same in this model. Although the maximum friction coefficient in rocks is generally known to be around 0.6 \cite{tullis2007friction, Byerlee}, an overestimation of friction was avoided by uniformly using a friction coefficient = 0.3.
Rock expansion due to irradiation degradation was evaluated by performing irradiation expansion calculations using the above parameters for model.

\begin{figure}
\includegraphics[width=\linewidth]{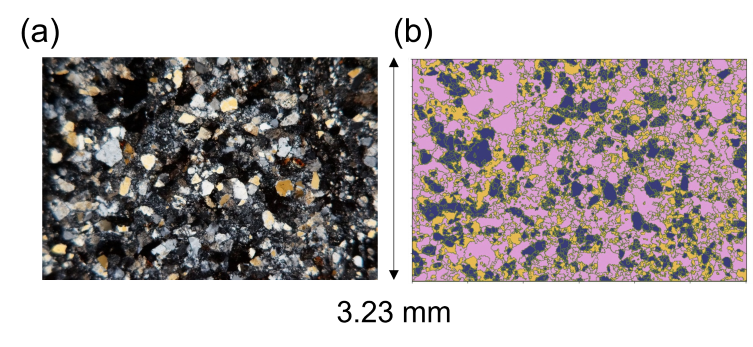}
\caption{
(a) Polarization microscope image of sandstone. 
(b) Results of mineral and grain boundary identification by image analysis. Pink indicates quartz, yellow indicates feldspar, and dark blue indicates biotite.
}
\label{fig:stream}
\end{figure}

\begin{table}[]
\caption{
Percentage of minerals composing the rock. 
}
\begin{tabular}{cccc}
         & Sandstone & Tuff    & Granite \\
   \hline   \hline
Quartz   & 58.3 \%   & 88.2 \% & 34.1 \% \\
Feldspar & 23.4 \%   &  -       & 57.6 \% \\
Chlorite & 18.3 \%   & 3.9 \%  &    -     \\
Biotite  &  -         & 7.9 \%  & 8.3 \%  \\
\end{tabular}
\end{table}

\section{Result and Discussion}
Table 2 shows the fracture stress parameters in the FEM that well explain the experimental values of tensile strength in the sandstone and tuff, with errors of 6\% and 1\%, respectively \cite{02bussei,03bussei}. 
The tensile strength of the granite was not reproduced by the analysis even by using the reported experimental values of fracture stress, and the tensile strength in the FEM model showed excessive values even when the friction coefficient was set to as small as 0.001.

\begin{table}[]
\caption{
FEM parameters of rocks obtained from Brazilian tests. 
}
\begin{tabular}{cccc}
                       & Sandstone       & Tuff            & Granite        \\
                          \hline   \hline
Fracture stress (MPa)  & 12              & 20              & 5              \\
Friction coefficient   & 0.3             & 0.3             & 0.001          \\
Reaction force (kN)    & 0.0031          & 0.0033          & 0.00915        \\
%Tensile strength (MPa) & 19.1 (+6.11 \%) & 20.3 (-0.98 \%) & 56.3 (+766 \%) \\
Tensile strength (MPa) & 19.1 & 20.3 & 56.3 \\
\end{tabular}
\end{table}

To investigate the tendency of failure in the Brazilian test, the principal stresses for each rock were examined when the reaction force reached its maximum value in the Brazilian test. Principal stress contour plots for each rock are shown in Fig. 2. 
In the tuff with the smallest error in tensile strength, the stresses are linearly generated from the point of forced displacement to the point of constraint. On the other hand, in granite, the stress is anisotropic and excessive stress is generated in the mineral grains of the granite. This excess stress is expected to be resolved by fracture within the mineral grains of the granite, which is not originally assumed in this FEM model. Alternatively, it could be resolved by performing the analysis for even larger granites and creating a model in which a grain boundary appears between the forced displacement point and the constraint point. However, the actual size of the mineral grains in the experiment is unknown, and it is also unknown whether fracture occurred inside the mineral grains in the experiment. The results of this FEM analysis indicate that the relationship between rock grain size and tensile strength should be carefully investigated in experiments.

\begin{figure*}
\includegraphics[width=\linewidth]{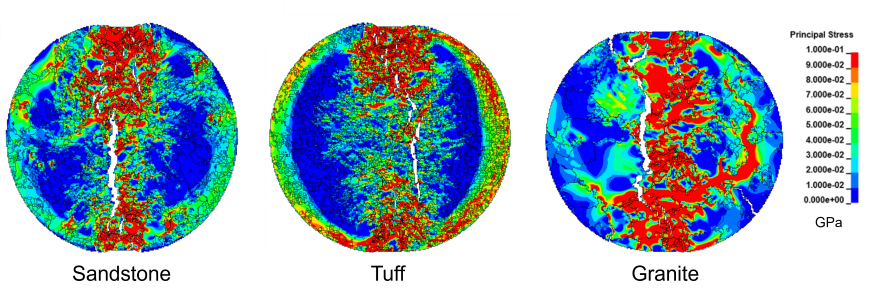}
\caption{
Contour plot of maximum stress during Brazilian test. 
}
\label{fig:stream}
\end{figure*}

We treated irradiation accumulation as a thermal expansion phenomenon in a sufficiently thermal equilibrium state, and analyzed the irradiation expansion phenomenon by FEM using the above parameters. The volumetric expansion rate $\Delta V$ and the linear strain $\varepsilon$ of the rock resulting from irradiation accumulation are defined as follows: 
\begin{eqnarray}
\Delta V & = \frac{3}{2}\{ \left(1+\varepsilon\right)^2 -1 \}+1,
\varepsilon & = \frac{{\mathrm{\Delta l}}_{ave}}{\sqrt{\left(\frac{X}{2}\right)^2+\left(\frac{Y}{2}\right)^2}},
\end{eqnarray}
where ${\mathrm{\Delta l}}_{ave}$ is the average displacement of the nodes at each vertex of the FEM mesh, X is the length of the analytical model in the $x-$direction, and Y is the length in the $y-$direction. 
The maximum, minimum, and average values of the volumetric expansion of each rock for each neutron fluence are calculated and are shown in Fig 3.  Experimental values of the irradiation expansion for sandstone and granite have been reported by Denisov et al. \cite{Denisov}, and our FEM calculations reproduce the absolute values and qualitative trends well. 

According to Denisov et al.\cite{Denisov}, for quartz-rich rocks in general, the trend of irradiation expansion can be expressed by the following: 
\begin{eqnarray}
\frac{\mathrm{\Delta V}}{V}=\frac{a{\frac{\mathrm{\Delta V}}{V}}_{max}\left(e^{bn_{cm}}-1\right)}{{\frac{\mathrm{\Delta V}}{V}}_{max}+ae^{bn_{cm}}}, 
\end{eqnarray}
where $a$ and $b$ are parameters that depend on the mineral species, $\left(\mathrm{\Delta V}/V\right)_{max}$ is the maximum expansion rate of the mineral, and $n_{cm}$ is the number of displaced atoms, proportional to the irradiation dose.
This function shows an S-shaped curve, which is consistent with the irradiation expansion results shown in the Fig. 3.

\begin{figure*}
\includegraphics[width=\linewidth]{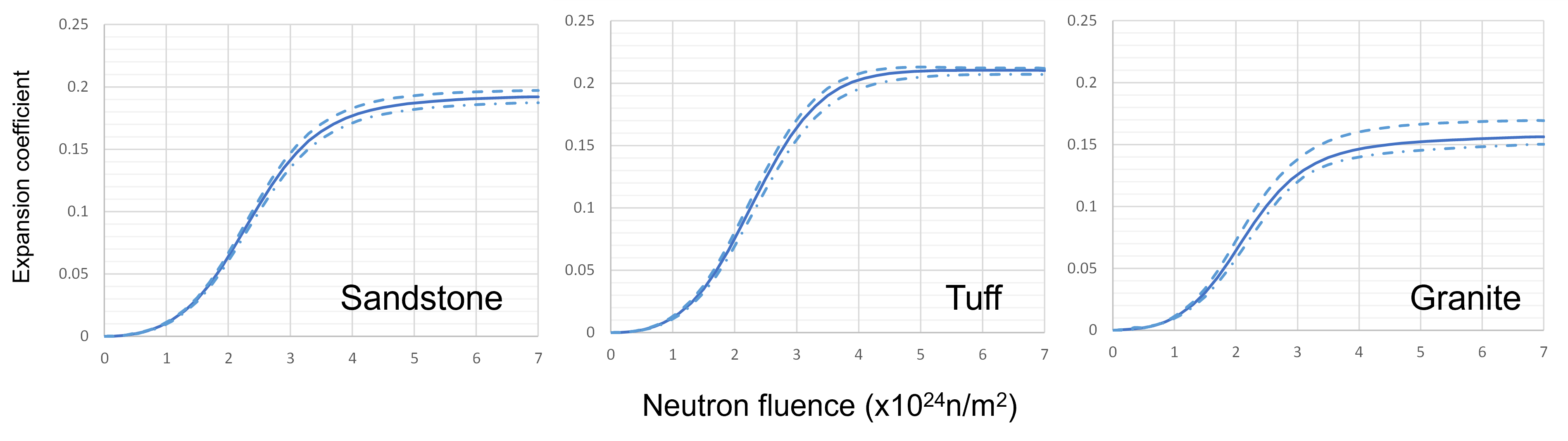}
\caption{
Results of FEM analysis of volumetric expansion rate with irradiation accumulation in rocks. The solid line indicates the mean value, and the dashed lines indicate the maximum and minimum values, respectively.
}
\label{fig:stream}
\end{figure*}

The maximum volumetric expansion of individual minerals calculated from the linear expansion coefficients used in our input is 17.9\% for quartz \cite{Denisov}. 
The maximum expansion is higher than that of quartz in all the rocks analyzed in this study. 
These results suggest that the volumetric expansion of rocks due to radiation accumulation is not only caused by linear expansion of minerals, but also by fracture at grain boundaries between minerals, which creates voids that increase the volumetric expansion of rocks. 
In fact, the present analysis shows that fracture occurs in the FEM mesh at an irradiation dose of $1\times10^{24} neutron/m^2$. In addition, the maximum expansion of the rock tends to be higher for rocks with higher quartz content in the range analyzed in this study.

In order to understand the phenomenon of irradiation expansion in rocks, stress contour plots were output for each neutron fluence. 
The stress contour diagram for tuff is shown in Fig. 4. 
Note that, stress contour images for sandstone and granite are similar to those for tuff. 
From the stress contour diagram, it can be seen that at neutron fluence $0.5\times10^{24} neutron/m^2$, a localized spot of fracture stress is observed in the interior of the rock, which tends to spread over the entire rock from fluence $1-3\times10^{24} neutron/m^2$. 
The irradiation doses from $1-3\times10^{24} neutron/m^2$ correspond to an increase in the linear expansion coefficient of quartz. 
Although the coefficient of linear expansion of quartz decreases and saturates after $3\times10^{24} neutron/m^2$, the expansion of the rock continues to increase. This may be attributed to the fact that the fracture of the rock allows voids to be generated at the grain boundary, and these voids gradually develop. Based on the above results, the following three-step model can be considered as the irradiation expansion mechanism of the rock: 
(1) $\sim 0.5\times10^{24} neutron/m^2$: irradiation expansion due to the superposition of expansion of each mineral that does not cause fracture inside the rock,
(2) $1-3 \times10^{24} neutron/m^2$: irradiation expansion with fracture inside the rock due to rapid expansion of quartz,
(3) over $3 \times10^{24} neutron/m^2$: irradiation swelling due to the generation of cracks.
%%%%%%%%%%
The result of our FEM analysis gives a reasonable interpretation of the curve drawn by functional of Eq. 2.
%%%%%%%%%%

\begin{figure}
\includegraphics[width=\linewidth]{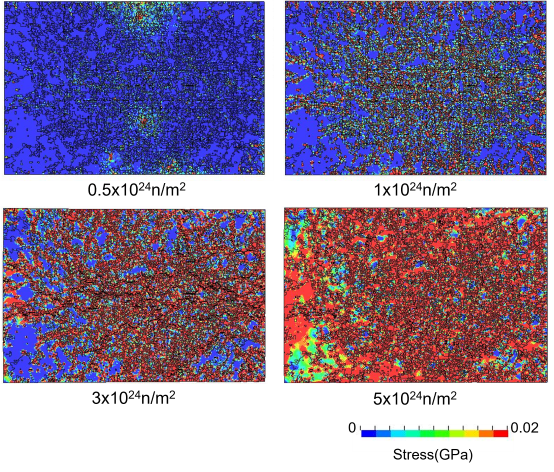}
\caption{
Dependence of irradiation accumulation on stresses inside tuff obtained from FEM analysis.
}
\label{fig:stream}
\end{figure}

Our present FEM model does not take into account the fracture inside the mineral, as mentioned above. The fracture of the mineral interior is expected to affect the generation and relaxation of local stress, especially in low-irradiation regions, as shown in the stress contour diagram (Fig. 4). Therefore, with further detailed experiments, a more accurate FEM model will be considered by modeling the fracture phenomena associated with irradiation accumulation based on the experimental results, especially in the low-irradiation regions. %Anyway, it can be said that the current FEM model already adequately reproduces the phenomena of irradiation degradation in the large irradiation accumulation region.

\section{conclusion}

Our present FEM modeling shows that it is possible to reproduce the fracture and expansion phenomena associated with irradiation accumulation of rocks used in concrete buildings in large radiation accumulation regions through a two-dimensional FEM analysis using experimental values of linear expansion coefficients of minerals as inputs. 
Using this model, it will be possible to clarify radiation-induced degradation phenomena for a wide variety of rocks from the viewpoint of computer science. The present analysis also indicates that careful investigation of fracture phenomena inside minerals is necessary, especially in mechanical tests and irradiation experiments in low-irradiation regions. 
Clarifying the fracture phenomena in regions of small irradiation accumulation in rocks are important from the viewpoint of maintenance of nuclear power plant. It is expected that our model will be refined by feedback from experiments, and that the safety of nuclear power plants will be discussed from computer simulations in the future.

\section*{acknowledgments}
This work constitutes part of the Japan Concrete Aging Management Program on Irradiation Effects (JCAMP), sponsored by METI in Japan.

\bibliography{apssamp}% Produces the bibliography via BibTeX.

\end{document}